Historical Patterns and Recent Impacts of Chinese Investors on United States Real Estate

Kevin Sun

Unionville High School

**Abstract**

Since supplanting Canada in 2014, Chinese investors have been the lead foreign buyers of U.S. real estate, concentrating their purchases in urban areas with higher Chinese populations like California. The reasons for investment include prestige, freedom from capital confiscation, and safe, diversified opportunities from abroad simply being more lucrative and available than in their home country, where the market is eroding. Interestingly, since 2019, Chinese investors have sold a net 23.6 billion dollars of U.S. commercial real estate, a stark contrast to past acquisitions between 2013 to 2018 where they were net buyers of almost 52 billion dollars worth of properties. A similar trend appears in the residential real estate segment too. In both 2017 and 2018, Chinese buyers purchased over 40,000 U.S. residential properties which were halved in 2019 and steadily declined to only 6,700 in the past year. This turnaround in Chinese investment can be attributed to a deteriorating relationship between the U.S. and China during the Trump Presidency, financial distress in China, and new Chinese government regulations prohibiting outbound investments. Additionally, while Chinese investment is a small share of U.S. real estate (~1.5% at its peak), it has outsized impacts on market valuations of home prices in U.S. zip codes with higher populations of foreign-born Chinese, increasing property prices and exacerbating the issue of housing affordability in these areas. This paper investigates the rapid growth and decline of Chinese investment in U.S. real estate and its effect on U.S. home prices in certain demographics.

**Historical Patterns and Recent Impacts of Chinese Investors on U.S. Real Estate**

**Introduction**

\*In this paper, the term "Chinese investors" is solely used to describe investors from China\*

Today, many of the largest cities in the United States face housing affordability crises, as a relatively stagnant housing supply fails to uphold the high demand of people wanting to live in urban cities. Simultaneously, there has been a substantial influx of wealthy home buyers from overseas investing in the domestic housing markets. Specifically, this paper outlines the behavior of Chinese buyers in the U.S. real estate market. This demographic has been leading foreign investments in U.S. homes for the past nine years and the capital flow from China to the U.S. has shown substantial impacts on American consumers by increasing housing prices in certain locations and zip codes. The following content of this research paper examines the existing surrounding theories, problems, and impacts of Chinese investors in U.S. real estate.

*Chinese Investor Profile*

First, when considering the typical Chinese investor profile, the dominant players are state-owned enterprises, sovereign wealth funds, banks, and high-net-worth private investors. In particular, Chinese banks and sovereign wealth funds show a sincere interest in the U.S. commercial real estate sector. For instance, from 2005 to 2014, the Bank of China made over 35 transactions in U.S. CRE, amassing over $8.4 billion in refinancing and sales, a huge figure in its timeframe (Mahajan & Sheth, 2014). Likewise, the Chinese Investment Corporation, the largest sovereign wealth fund in the world with more than $1.3 trillion in assets around the globe, is an active participant in U.S. CRE through lending and fund vehicles. However, while these original investors were the makeup and composition of nearly all Chinese investing in the early 2010s, the investor profile has diversified since and expanded beyond these original entities with Chinese regulators opening up international investments to other financial institutions like insurance companies.

**Background**

Up until 2010, direct Chinese investment in the U.S. real estate market had been fairly negligible (Ehrlich, 2022). Yet, within the last decade, Chinese investors have risen to become the lead foreign buyers of U.S. homes, spurring the question of what drew them to the United States in particular, especially with similar opportunities available in western countries such as the United Kingdom (Olick 2021). Furthermore, while Chinese overall investment in foreign real estate assets has increased across all countries, this decade introduced a major shift in focus to U.S. investment specifically, from treasury bonds to stocks to venture capital, with real estate having seen some of the most significant growth (Fung, 2019). Although there is a multitude of reasons for the redirection of resources to the U.S., the most attractive features of investing in real estate in the U.S. are detailed below.

*Reasons for Investing*

One of the motivating factors behind investing in the United States is that it is home to about three-quarters of the top schools in the world. Because education is deeply ingrained in traditional Chinese culture, the best universities are highly desirable. In fact, many wealthy Chinese parents seek a second home in the U.S. so that their children can establish residency to have easier access in attending elite American colleges. According to a 2012 Hurun Report, one in four Chinese nationals who are worth more than $16 million have emigrated for this reason

with another 47 percent considering emigration (Carlson, 2012). A hotspot for this type of migration is California, where state legislation requires a person to live in the state for only 366 days before they can establish residency and receive benefits for college such as in-state tuition, a nearby home, and higher acceptance rates among the prestigious public universities like the University of California, Berkeley or the University of California, Los Angeles, the number one and two ranked public universities according to U.S. News ("Top Public", 2022).

Even historically, Chinese home buyers have largely targeted their efforts on the U.S. coasts such as New York and California (both culturally friendly states). With almost 40% of Chinese real estate investments concentrated in California and over a billion dollars invested from 2005 to 2014 (see Fig. 1), California is a converging hotspot due to its thriving Chinese community, diversified economy, and plentiful prestigious academic universities (Passy, 2019). Recently, however, a new wave of Chinese investors has emerged, looking for real estate opportunities in areas that were never sought after before, such as properties in the Midwest and Southwest which are home to distinguished institutions such as the University of Chicago ("5 Reasons", 2022).

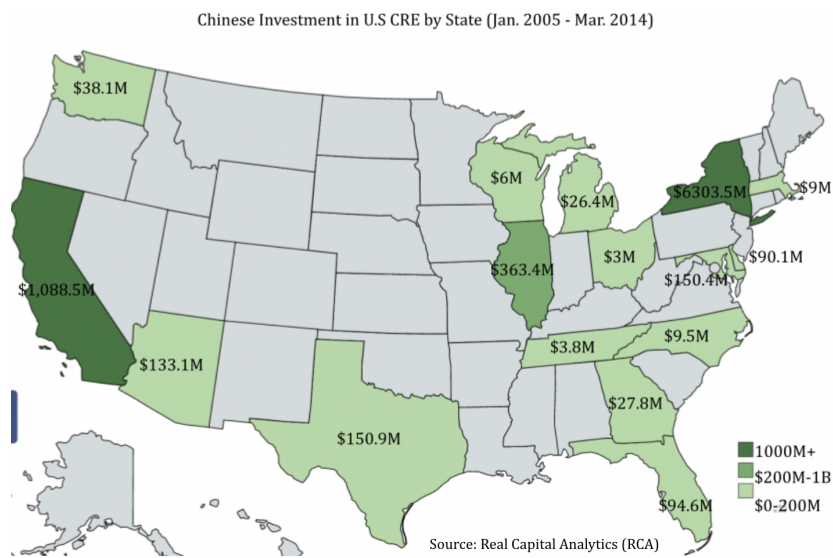

Fig. 1: Chinese investment in U.S. CRE by State. Data from Real Capital Analytics (RCA)

In addition to the vast educational opportunities, an increase in demand for residential real estate by Chinese buyers can also be attributed real estate investing serving as a viable option for citizenship. More specifically, the EB-5 visa can be used as a green card for wealthy

Chinese investors, allowing for a direct path to U.S. citizenship. Under this program, investors (and their spouses and children under 21) are eligible to apply for a Green Card and permanent residence if they make the necessary minimum investment of $1,050,000 in a commercial enterprise in the United States. Furthermore, a 2014 survey conducted by Hurun, a Shanghai research firm, revealed that approximately 64 percent of Chinese individuals with assets of more than $1.6 million were either emigrating or planning to do so. By replacing their Chinese assets with U.S. assets, these individuals can obtain citizenship in just six months with the EB-5 visa, further highlighting the attractiveness of U.S. citizenship through real estate investing ("Hurun Report Chinese," 2014).

      Perhaps most important, U.S. real estate is a much more enticing, lucrative, and safer vehicle to invest in than mainland China. Because Chinese investors primarily purchase in urban cities, many Chinese and Hong Kong nationals will choose the U.S. because of its cheaper housing prices compared to the metro areas of other western countries. Foreign investments also allow for the diversification of assets and avoiding domestic taxes. As for the wealthy Chinese billionaires, U.S. real estate serves as an outlet to appreciate their already-attained wealth. From data about foreign tax havens, the International Consortium of Investigative Journalists reported in 2014 that over 22,000 clients of offshore financial institutions had addresses in China and Hong Kong, including some of China's most powerful, rich, and influential men and women (Walker et al., 2014). The U.S. is particularly desirable for these rich individuals because of its political stability and the fact that it is one of the fairest and most just countries in the world (Dogen, 2022). Because of this, the U.S. serves as a safer place to invest in comparison to China's sudden-changing policies and volatile environment for investing. For instance, with China's one-party political system, Chinese money is susceptible to new policies confiscating

their wealth. Investors abroad are looking to protect their wealth from the volatility at home and the U.S. market provides that reassurance. After all, investors need to feel financially secure to feel rich.

Prestige is another alluring feature evident through past Chinese acquisitions. In 2015, Anbang Insurance Group Co. paid the highest purchase price ever for a U.S. hotel with its 1.95 billion dollar purchase of the Walford Astoria in New York (Fung, 2019). By investing in blockbuster deals and acquiring trophy buildings, investors are able to strengthen China's prestige and reputation. Other landmarks like the Vista Tower, a nearly $1 billion skyscraper in Chicago developed by China's largest commercial property company, Dalian Wanda, and an eight-acre luxurious housing development project in Beverly Hills are clear demonstrations of this initiative (Fung, 2019).

Finally, investing across the world in a completely different country is a learning process for Chinese investors. According to a staff report by the U.S.-China Economic and Security Review Commission, the acquisition of assets like office buildings and hotels that require a long-term commitment for rental incomes is viewed as an opportunity to become familiar with the local tax system, and a foundation for further developments in the future (Koch-Weser & Ditz, 2015). These larger Chinese investors seek to better understand local markets before expanding to high-stakes ventures such as greenfield projects and billion-dollar development projects.

**The Prosperity Era**

The early 2010s have proved to be an extremely successful period for Chinese investors in both commercial and residential U.S. real estate. In 2014, China supplanted Canada to become the lead foreign investor in the U.S. real estate market. In the residential sector, from 2013 to 2018, the total number of properties purchased by the Chinese nearly doubled from 23,100 to 40,400 (see Fig. 2: Statistia Research Department, 2022). Additionally, from 2011 to 2016, Chinese nationals purchased homes valued at $93 billion, including $28.6 billion in 2015 alone ("Breaking Ground," 2016). From the commercial perspective, these investors have shared an equally impressive growth where between 2013 to 2018 they were net buyers of almost 52 billion dollars worth of properties (Putzier, 2022).

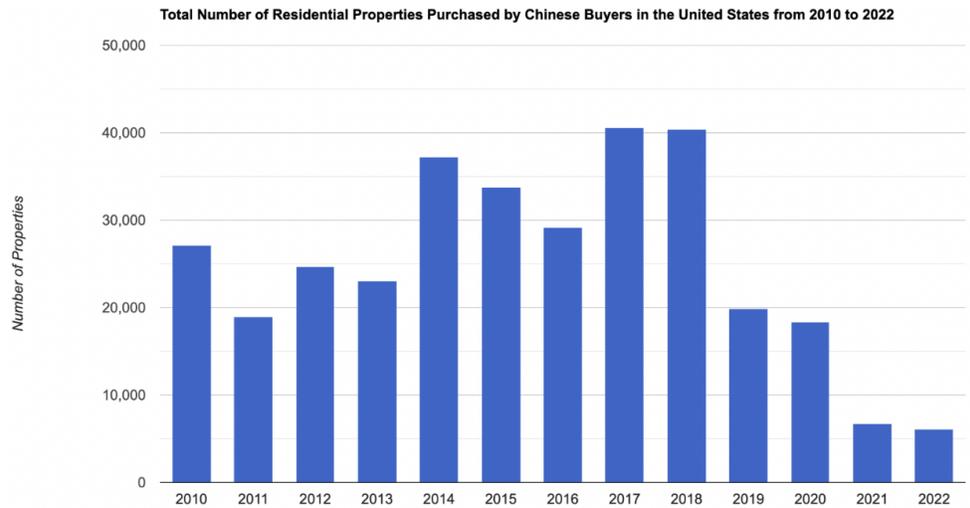

Fig 2: Total Number of Residential Properties Purchased by Chinese Buyers in the United States from 2010-2022 (Data from Statistica)

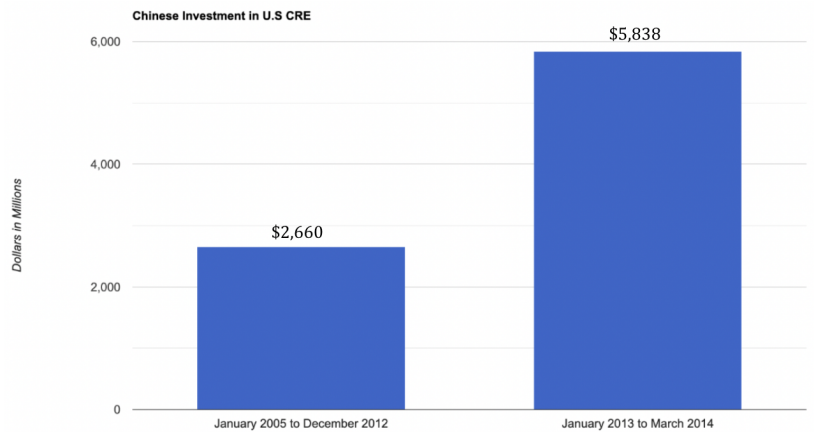

Fig 3: Chinese Investment in U.S. CRE. Source: Real Capital Analytics

With a peak in commercial investments at $9.3 billion in 2016, the pinnacle represents a 15-fold increase from just six years ago in 2010 ("Breaking Ground," 2016). In addition, at approximately $208 billion, China was the biggest foreign holder of U.S.-government-backed real estate bonds in 2016 ("Breaking Ground," 2016).

Besides the draws of education, prestige, citizenship, and safe yields in U.S. real estate, an important factor leading to this era of prosperity was the abundance of lucrative investments available after the 2008 housing crash. The real estate market following the crash of 2007-2008 led to spectacular opportunities for individuals with knowledge and access to capital. Higher inventory and less demand caused deflated property prices, which eventually lead to greater long-term returns and house appreciations. For example, the average home valuation fell to $257,000 in 2009, but with an average investor hold of 5 years, the average home valuation rebounded and jumped to $369,400 ("Average Sales", 2022). Experienced investors could invest in a cheap and ripe market and receive a 43% return on investment. Additionally, according to the latest report from Cohen & Steers, an investment management company with over $56 billion invested in real estate, "superior returns in real estate tend to follow recessionary periods" (Zhang, 2022). Another reason for this spike in investment is that, until 2012, the Chinese government prohibited insurance companies from buying foreign property. With these restrictions uplifted, the Chinese seized their invitation to venture into investments abroad, flooding Chinese money into U.S. real estate. The Wall Street Journal classified this period as "the greatest episode of capital flight in history" (Brown, 2016). This pattern of investment would persist, though not as rapidly, until the end of 2018.

      The next year was the first instance where the U.S. real estate market saw a reversal in behavior and a withdrawal of Chinese investment. Instead of continuing as net buyers of real estate, investors began selling off their U.S. properties. And although coinciding with the inception of the coronavirus, surrounding data and statistics reveal that the pandemic was not the primary factor for this turnaround, at least not directly.

**The Fall**

The withdrawal of Chinese investors in U.S. real estate was largely unexpected. Speculators and industry professionals strongly believed in 2016 that investment would only continue to flourish to close the decade (Grant, 2016). However, after 2018, purchases dropped significantly and Chinese investors began to sell more in property valuation than they bought, becoming net sellers for the first time in the 2010s. Also in 2018, the Beverly Hills Project bought by the Dalian Wanda Group was sold for only 420 million to a London-based real-estate firm Cain International and Alagem Capital Group, the exact price the conglomerate paid for the development back in 2014. Their eagerness to leave the U.S. CRE market was a foreshadowing of future Chinese companies bailing on U.S real estate after years of acquiring these properties. According to MSCI Real Assets, these investors sold a net 23.6 billion dollars of U.S. CRE since 2019, while between 2013 to 2018, they were net buyers of almost 52 billion dollars of U.S. commercial properties (Putzier, 2022).

Similarly, a weaker appetite is apparent in residential real estate as well. Research released by the National Association of Realtors found that the Chinese purchased $17 billion less on homes in America in 2019, a staggering 56 percent decline within 12 months (see Fig. 4).

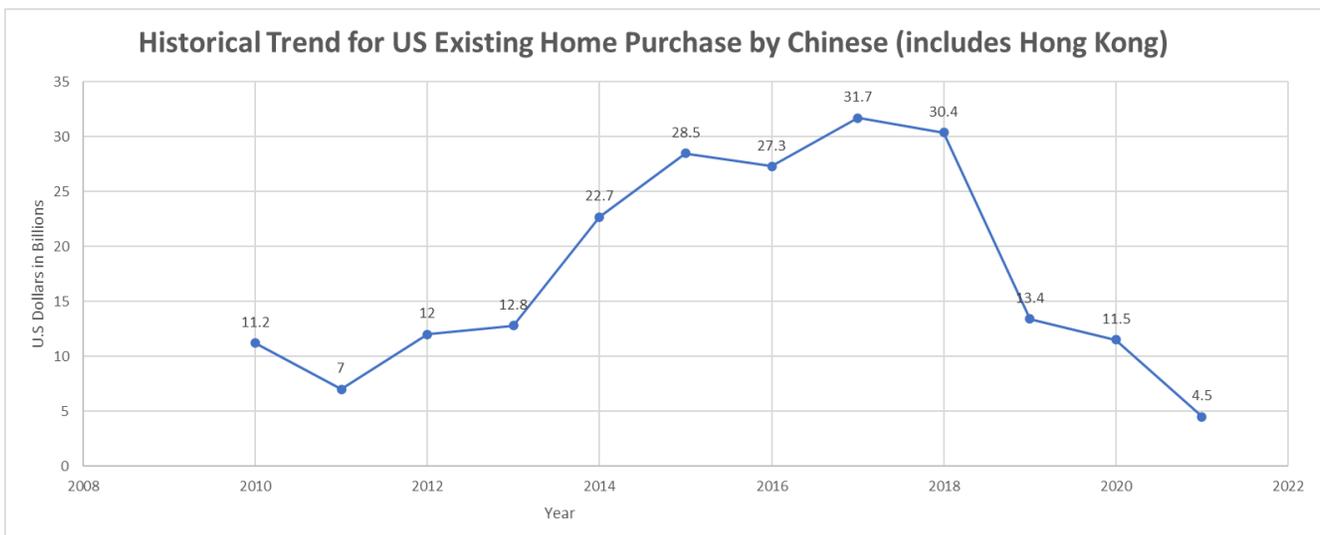

Fig 4: Historical Trend for U.S. Existing Home Purchases by Chinese Investors (Residential)
Source of Data: National Association of Realtors (NAR)

Additionally, in a short two-year window, housing purchases fell from nearly $32 billion in 2016-2017 to only $11.5 billion in 2019 (Keys, 2020). Even as recently as last year, Chinese buyers only spent a mere $6.1 billion total on both sectors of U.S. real estate, a fraction of what it had been during the thriving expansion of the early 2010s (Zilber, 2022).

**Causes of Withdraw and Selling Behavior**

With the pandemic's inception in 2019, many predictions about real estate went awry. However, the COVID-19 pandemic may not have been as impactful as many people assume for Chinese investors beginning to sell off their real estate. The strongest influences for this shifting behavior are actually restrictive government regulations, a deteriorating relationship with the U.S. during the trade war, and financial distress for large Chinese corporations and conglomerates.

  To begin, in 2016, the Chinese government placed additional hurdles on its citizens. China began restricting outbound investments, allowing residents to take only $50,000 out of the country in an attempt to raise the value of the Renminbi, the nation's currency. Although Chinese investors were still net buyers at this time, 2016 marked a peak in purchases with the introduction of legislation like this that prompted years of declining investment in the U.S. that hasn't happened since the early 2000s. According to data from Thomson Reuters, China's outbound M&A volumes nearly halved in the first six months of 2017 to $64.2 billion following the crackdown on capital outflows, after Chinese companies spent a record $221 billion on assets overseas in 2016 (Wu, 2017). Also in 2017, the General Office of the State Council of China released the "Guiding Opinions for Further Guiding and Regulating the Direction of Outbound Investments" in which they stated they would prohibit "irrational" investments including foreign real estate and ban domestic enterprises from participating in outbound investments that

"endanger or may endanger national interests or national security" ("China's New Policy", 2017). Essentially, the Chinese government believed that investment corporations were overpaying for assets abroad and taking on too much risk, prompting many companies to sell their assets and holdings. Ultimately, efforts from the Chinese government to stabilize its currency, reduce debt, and alleviate the country's economic stagnation have made it much more difficult to purchase real estate in America.

      The drastic shift from buying to selling is also caused by escalating political tensions between the U.S. and China. A report from the New York Times states, "growing distrust between the United States and China has slowed the once steady flow of Chinese cash into America, with Chinese investment plummeting by nearly 90 percent since President Trump took office" and in the period 2017-2019 (Rappeport, 2019). The falloff of investment stems from increased regulatory scrutiny by the U.S. and a colder attitude toward Chinese investment overall. For instance, Neil Brookes, Asia Pacific head of capital partners at Knight Frank, explained that Chinese outbound capital fell 83% in 12 months from 2018-2019 "largely due to trade wars and the government trying to stop money leaving the country" (Shao, 2019). The deterioration of their economic relationship during the trade war has scared businesses in both countries. From the U.S. perspective, a September 2019 study by Moody's Analytics found that the trade war had already cost the U.S. economy nearly 300,000 jobs and an estimated 0.3-0.7% of real GDP (Hass & Denmark, 2020). A 2019 report from Bloomberg Economics estimated that the trade war cost the U.S. economy $316 billion at the end of 2020, while more recent research from the Federal Reserve Bank of New York and Columbia University found that U.S. companies lost at least $1.7 trillion in the price of their stocks as a result of U.S. tariffs imposed on imports from China (Hass & Denmark, 2020). China also felt economic pain as a result of the

trade war with a slowdown in industrial output and economic growth. Furthermore, as the trade war dragged on, Beijing lowered its tariffs for its other trading partners as it reduced its reliance on U.S. markets (Hass & Denmark, 2020). With efforts to circumvent one another, it is unsurprising that Chinese buyer inquiries on U.S. property were down 27.5% from the previous year and up in Canada, the UK, Australia, and Japan, all of which served as alternative destinations to invest in real estate to the United States (Shao, 2019). Finally, additional tightening of borders and restrictions on immigration have made U.S. real estate less hospitable overall for Chinese investors. A May report from Cushman & Wakefield noted that in 2018, there were 37 property acquisitions by Chinese buyers worth $2.3 billion, but $3.1 billion of commercial real estate was sold off (Rappeport, 2019). The report said that the treatment of the Chinese conglomerate HNA Group and the tough trade talk made Chinese investors feel unwelcome and discouraged from investing (Rappeport, 2019).

The final reason for the withdrawal was that many Chinese investment companies were in financial distress and needed quick capital which they could easily acquire by selling their foreign assets. The timing was ripe for selling too. Investors who bought U.S. commercial properties a decade ago were still able to make profits by selling in 2019. In conjunction with a decline in business travel and a weak demand for office buildings, the beginning of the pandemic presented the perfect opportunity to sell. But ultimately, it was the combination of all of these factors that led to a sharp decline in Chinese investment in U.S. real estate, a trend depicted in Figure 5.

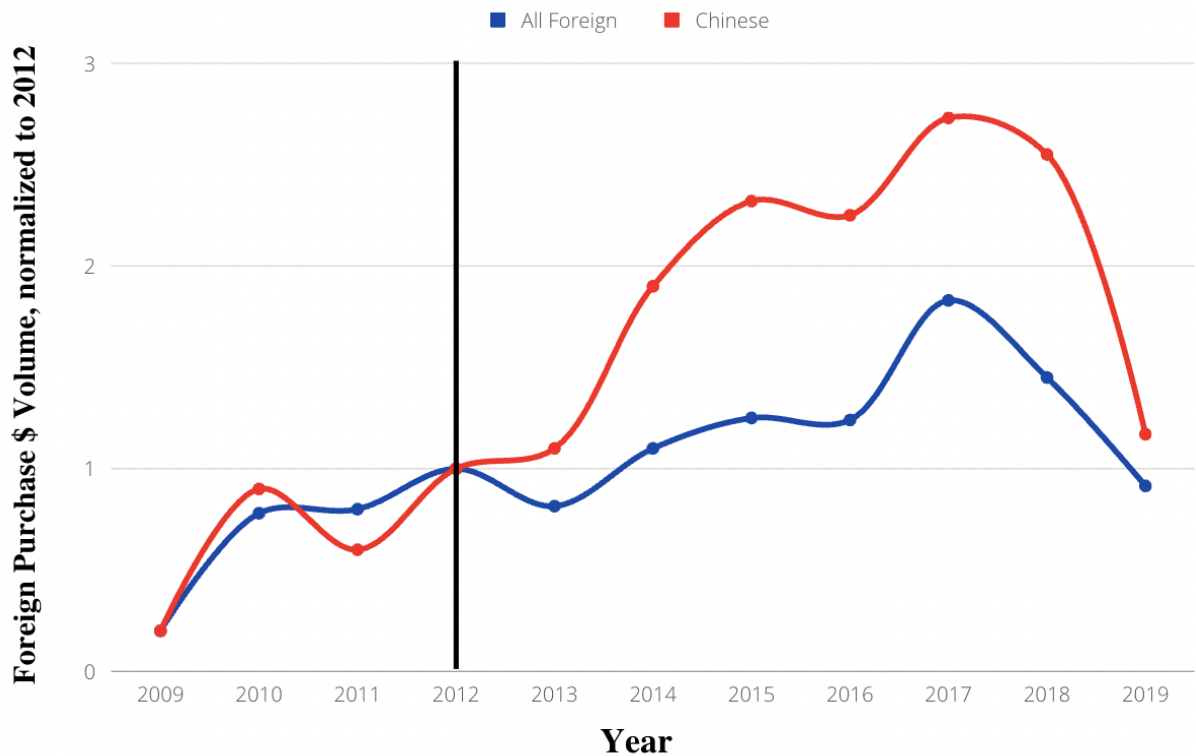

Fig 5: Sales Volume of Chinese vs. All Foreign Investors in U.S. Real Estate from 2009-2019, Normalized to 2012. Source: National Association of Realtors (NAR)

**Implications of Recent Behavior and Impact on Certain U.S. Homes**

So far, this paper has outlined the motivations and reasons for the turnaround of Chinese investment and the appealing factors for why they chose to invest in the U.S. initially. However, it is equally as important to acknowledge the implications of their recent behavior on the prices of U.S. properties. Essentially, what was the effect of the flooding of Chinese money on the U.S. real estate sector? Interestingly, while Chinese money always was a really small share of the U.S. real estate market even at its peak (around 1.5%), it had outsized impacts on price valuations in large U.S. cities and in the popular sites of California ("International Transactions", 2022). Because Chinese companies would often buy the most expensive hotels and office buildings in the most expensive cities (like New York and Los Angeles), they would pay incredibly high prices which then became benchmarks for other buildings. In this way, Chinese money had inflated real estate values for a long time. With its removal, the prices of surrounding buildings have fallen back down ("Spotlight on China", 2022).

Additionally, research from Wharton professor Benjamin Keys suggests that Chinese investors also have a significant impact on housing prices in locations with high foreign-born Chinese ethnic people. According to the findings in their research paper titled "Global Capital and local Assets: House Prices, Quantities, and Elasticities", housing prices grew 8 percent more in zip codes with high foreign-born Chinese populations from 2012 to 2018 (Gorback & Keys, 2020). Another study titled "Capital Flows, Asset Prices, and the Real Economy: A "China Shock" in the U.S. Real Estate Market" further supports the notion that foreign Chinese investments have a significant effect on local housing and labor markets. The study found that a one standard deviation increase in exposure to these purchases explains 24% and 18% of the cross-ZIP-code variation in local house prices, exacerbating the current worries about housing affordability and driving out lower-income residents (Li et al., 2020). Additionally, a 1% increase in housing demand by foreign Chinese, increased local home prices by 0.099% and 0.173%, respectively (Li et al., 2020). In regards to driving out local low-income residents, they found a slightly negative relationship between foreign Chinese housing transactions and the number of tax filings, suggesting that foreign Chinese house purchases drive out local residents on the net. They concluded that a 1% increase in foreign Chinese housing demand, as measured by transaction value and count, decreases the number of tax filings by 0.04% and 0.07%, respectively (Li et al., 2020). Finally, research conducted by Iacob Koch-Weser and Garland Ditz found that by buying homes chiefly for investment purposes, Chinese buyers can worsen housing bubbles (Koch-Weser & Ditz, 2015). For example, in San Francisco, the real estate cycles are between five to seven years. The current increase in prices from Chinese investment activity that are only three years old could last for another several years, again making housing less affordable for lower-income citizens in the area (Koch-Weser & Ditz, 2015).

**Future and Long-Term Effects**

While investment from China has rapidly declined in recent years, the outlook for the U.S. real estate market in 2022 is still very positive. Because Chinese money encompasses such a small portion of U.S. real estate, it has minimal influence on the sector as a whole. Demand remains strong in this inflationary environment, which commercial real estate has the ability to hedge against. Furthermore, a retreat by Chinese buyers could actually be good news for Americans looking to purchase a home, especially in the expensive markets in California, where its popularity had garnered a double-digit property appreciation. Ultimately, the lack of competition from foreign buyers, who typically invest with competitive all-cash offers, has resulted in better deals on homes for local buyers. This effect has been exemplified in the first half of 2019 in Irvine, where the median home sale price fell from $834,000 to $820,000 according to Zillow (Zhang, 2019). Additionally, in recent years leading up to 2019, Chinese investors made up about half of all home purchases in the city, but that share has fallen to about 36% in 2019 (Zhang, 2019). Although not solely the cause, the withdrawal of Chinese investment in Irvine was an important factor contributing to the drop in median home sale prices in California where house appreciation is rapid and constant.

However, the future for Chinese investors is not quite as certain. Because of government policy reducing funding available to developers, China's housing market began experiencing a real estate crisis in 2021. The property slump has persisted for over a year due to a weak economy and strict COVID-19 regulations discouraging buyers. According to a survey in October 2022 by China Index Academy, a top real estate research firm, the sales of the 100 biggest Chinese real estate developers dropped by 43% since 2021 (He, 2022). Simultaneous in October, however, are recent indicators that point to a more positive turn of events. According to brokerage Cinda Securities Co. Ltd., China's housing market has shown signs of stabilizing (Cheng, 2022). The Chinese government financed more than 1 trillion RMB to promote investment and implemented regulations to remove administrative friction, encourage international investment, lower home purchase costs, and boost rational demand in China (Cheng, 2022).

**Conclusion**

This research paper outlined the two-sided story of Chinese investment in U.S. real estate. The real estate crash of 2008 was a warm welcome to foreign investment which the Chinese capitalized on, investing in search of political stability, diversification, and prestige. Then, following a half-decade window of prosperity, 2019 experienced a turnaround of behavior where investors became net sellers, a stark contrast to tens of billions of dollars in acquisitions between 2013 to 2018 (Statistia Research Department, 2022). This reversal is attributed to the combined factors of a deteriorating political relationship between the U.S. and China, financial distress in China, and government regulations on investment. And finally, this paper discussed the impacts of Chinese investment on certain U.S. properties. In major cities, landmark purchases from Chinese corporations inflated nearby office building valuations. In locations with high foreign-born Chinese populations, high degrees of foreign investment raised local housing prices and intensified concerns over housing affordability with already expensive rents in California. Should Chinese investment in U.S. real estate rebound, further research should investigate how effective government regulation can limit the inflow of Chinese capital, and thus reduce the volatility of housing prices in areas significantly impacted by Chinese investment.

HISTORICAL PATTERNS AND RECENT IMPACTS                                                           20Franklin Templeton Investments. (2022, January 5). *A Tale Of Two Real Estate Markets: U.S. And China*. SeekingAlpha.

https://seekingalpha.com/article/4478060-a-tale-of-two-real-estate-markets-us-and-china

*FRED Economic Data*. (2022, October 26). https://fred.stlouisfed.org/series/ASPUS

Fung, E. (2019, January 29). *Chinese Exiting U.S. Real Estate as Beijing Directs Money Back to Shore Up Economy*. WSJ.

https://www.wsj.com/articles/chinese-exiting-u-s-real-estate-as-beijing-directs-money-back-to-shore-up-economy-11548757800

Grant, P. (2016, May 25). *Chinese Investors Pour Money Into U.S. Property*. WSJ.

https://www.wsj.com/articles/chinese-investors-pour-money-into-u-s-property-1464110682

Guevara, M. W. (2021, July 9). *Leaked Records Reveal Offshore Holdings of China's Elite*. ICIJ.

https://www.icij.org/investigations/offshore/leaked-records-reveal-offshore-holdings-of-chinas-elite/

He, L. (2022, November 14). *China's real estate crisis could be over. Property stocks are soaring*. CNN.

https://edition.cnn.com/2022/11/14/investing/china-real-estate-crisis-over-rescue-plan-intl-hnk/index.html

*International Transactions in U.S. Residential*. (2017, July 18). www.nar.realtor.

https://www.nar.realtor/research-and-statistics/research-reports/international-transactions-in-u-s-residential-real-estate

Journal, T. W. S. (2022, September 21). *Chinese Firms Flee U.S. Commercial Real-Estate Market - The Wall Street Journal Google Your News Update - WSJ Podcasts*. WSJ.

HISTORICAL PATTERNS AND RECENT IMPACTS                                          21https://www.wsj.com/podcasts/google-news-update/chinese-firms-flee-us-commercial-real-estate-market/10ada6b2-b4d8-4baf-bf5d-371cfa2adb5c

Koch-Weser, I., & Ditz, G. (2015). *Chinese Investment in the United States: Recent Trends in Real Estate, Industry, and Investment Promotion*. CreateSpace Independent Publishing Platform.

Li, Zhimin, Leslie Sheng Shen and Calvin Zhang (2020). Capital Flows, Asset Prices, and the Real Economy: A "China Shock" in the U.S. Real Estate Market. International Finance Discussion Papers 1286.

Lueckemeyer, O. (2022, September 20). *Chinese Investors Sell Off Nearly >4B In U.S. Real Estate As Values Of Trophy Properties Fall*. Bisnow. https://www.bisnow.com/national/news/capital-markets/chinese-investors-sell-off-nearly-24b-worth-of-us-real-estate-as-values-of-trophy-properties-fall-115494

Mahajan, S., & Sheth, S. (2014). *Chinese investment in U.S. real estate Collaborate and benefit*. https://www2.deloitte.com/content/dam/Deloitte/us/Documents/financial-services/us-fsi-chinese-investment-in-US-real-estate-051614.pdf

Passy, J. (2019, May 16). *The Chinese purchase more U.S. residential real estate than buyers from any other foreign country, but Trump's trade war may change that*. MarketWatch. https://www.marketwatch.com/story/chinese-investors-buy-more-us-residential-real-estate-than-any-other-country-but-trumps-trade-war-could-soon-end-that-2019-05-15

Putzier, K. (2022, September 20). *Chinese Firms Flee U.S. Commercial Real-Estate Market After Big Property Bets Sour*. WSJ. https://www.wsj.com/articles/chinese-firms-flee-u-s-commercial-real-estate-market-after-big-property-bets-sour-11663622573